\documentclass{nature}

\usepackage{hyperref} 
\usepackage{graphicx}
\usepackage{amsmath}
\usepackage{amssymb}
\usepackage{siunitx}
\usepackage{braket}
\usepackage{bm}

\DeclareGraphicsExtensions{.png,.jpg,.eps}

\usepackage{xcolor}
\usepackage[normalem]{ulem}

\makeatletter
\let\saved@includegraphics\includegraphics
\AtBeginDocument{\let\includegraphics\saved@includegraphics}
\renewenvironment*{figure}{\@float{figure}}{\end@float}
\makeatother

\title{Topological gap solitons in a 1D non-Hermitian lattice}

\author{N.~Pernet$^{1}$, P.~St-Jean$^{1}$\footnote{These authors contributed equally: N.~Pernet, P.~St-Jean}, D.D.~Solnyshkov$^{2,3}$, G.~Malpuech$^2$, N. Carlon Zambon$^1$, B.~Real$^4$, O.~Jamadi$^4$, A.~Lema\^itre$^1$, M. Morassi$^1$, L.~Le~Gratiet$^1$, T.~Baptiste$^1$, A.~Harouri$^1$, I.~Sagnes$^1$, A.~Amo$^4$, S.~Ravets$^1$, J.~Bloch$^1$}

\begin{document}

\maketitle

\begin{affiliations}
\item Universit\'{e} Paris-Saclay, CNRS, Centre de Nanosciences et de Nanotechnologies (C2N), 91120, Palaiseau, France
\item Institut Pascal, PHOTON-N2, Universit\'e Clermont Auvergne, CNRS, SIGMA Clermont, F-63000 Clermont-Ferrand, France.
\item Institut Universitaire de France (IUF), 75231 Paris, France
\item Universit\'{e} de Lille, CNRS, Laboratoire de Physique des Lasers Atomes et Mol\'{e}cules (PhLAM), 59000 Lille, France
\end{affiliations}	

\begin{abstract}
Nonlinear topological photonics is an emerging field aiming at extending the fascinating properties of topological states to the realm where interactions between the system constituents cannot be neglected. Interactions can indeed trigger topological phase transitions, induce symmetry protection and robustness properties for the many-body system. Moreover when coupling to the environment \textit{via} drive and dissipation is also considered, novel collective phenomena are expected to emerge. Here, we report the nonlinear response of a polariton lattice implementing a non-Hermitian version of the Su-Schrieffer-Heeger model. We trigger the formation of solitons in the topological gap of the band structure, and show that these solitons demonstrate robust nonlinear properties with respect to defects, because of the underlying sub-lattice symmetry. Leveraging on the system non-Hermiticity, we engineer the drive phase pattern and unveil bulk solitons that have no counterpart in conservative systems. They are localized on a single sub-lattice with a spatial profile alike a topological edge state. Our results demonstrate a tool to stabilize the nonlinear response of driven dissipative topological systems, which may constitute a powerful resource for nonlinear topological photonics.
\end{abstract}
	
{
\let\clearpage\relax
\maketitle
}	

The topology of band structures in periodic systems is related to the existence of a non-zero Berry phase, and gives rise to fascinating phenomena like anomalous velocity, chiral edge states that are robust to disorder or topological Thouless pumping~\cite{Hasan2010,Qi2011,Kosterlitz2017,Haldane2017,Thouless1983}. First discovered in solid state systems, topological physics can also be emulated in artificial lattices, including atomic~\cite{Cooper2019}, photonic~\cite{Lu2014, Ozawa2019}, mechanical~\cite{Huber2016}, optomechanical~\cite{Peano2015}, and polaritonic~\cite{Gianfrate2020, Klembt2018} systems. These platforms have allowed engineering topological phases hardly achievable in condensed matter, involving synthetic dimensions~\cite{Ozawa2019a}, disorder effects~\cite{Meier2018, Stutzer2018}, quasi-crystalline structures~\cite{Kraus2012, Baboux2017, Dareau2017} or higher-order multipoles~\cite{Serra-Garcia2018, Peterson2018}. The physics becomes even richer when inter-particle interactions are considered. In the weakly interacting regime, nonlinearities can induce topological phase transitions~\cite{Hadad2016,Zhou2017}, or enable the formation of solitons in a topologically non trivial gap (thus called "topological gap solitons")\cite{Leykam2016, Solnyshkov2017,Mukherjee2020, Mukherjee2020_a,Xia2020a,Bisianov2019}. In the strongly interacting regime, novel symmetry protected phases may appear~\cite{DeLeseleuc2019}, and fractional quantum Hall physics can be emulated with the possible stabilization of multi-particles Laughlin states~\cite{ Clark2020}.

Recently, photonic platforms have allowed pushing this exploration beyond the realm of conservative Hamiltonians, mainly through the engineering of gain and losses. Non-Hermitian topological systems have led, for example, to the development of topological lasing~\cite{Solnyshkov2016, St-Jean2017, Zhao2018, Parto2018, Bahari2017, Bandres2018}, robust sources of non-classical light~\cite{Blanco-Redondo2018, Mittal2018, Kruk2019, Barik2018}, and PT-symmetric phases~\cite{Weimann2017, Song2019,Zeuner2015}. Most of these works on non-Hermitian topology have focused on probing, stabilizing or amplifying the linear response of the system. These recent advances now offer the possibility to experimentally explore topological photonics in a regime where non-Hermiticity and nonlinearity are combined~\cite{xia2020}.

In this article, we investigate the physics of topological gap solitons in a 1D driven-dissipative polariton lattice. We emulate a nonlinear and non-hermitian version of the well-known Su-Schrieffer-Heeger (SSH) model: a 1D bipartite topological lattice with staggered  hopping energies forming a chain of coupled dimers.~\cite{Su1979}. Polaritons are well suited for exploring topological photonics~\cite{St-Jean2017, Klembt2018, Whittaker2019,Gianfrate2020}. Indeed their excitonic fraction provides repulsive interactions, resulting in a Kerr-type nonlinearity, while their photonic component makes the system intrinsically non-Hermitian, with the balance of drive and dissipation playing a crucial role in its dynamics~\cite{Carusotto2013}.
	
We report the formation of solitons in the topological gap of the lattice and study their sub-lattice pseudospin properties. We investigate the robustness properties of these solitons using an optically controlled non-Hermitian defect, i.e. a local perturbation to the real and imaginary parts of the potential landscape. We demonstrate that these solitons are robust to defects located one sub-lattice, a property inherited from the chirality of the underlying model. The crucial novelty brought by the non-Hermitian character of the generated solitons appears when engineering the phase of the driving field. This engineering allows the onset of spin-polarized solitons that are not accessible in conservative systems. Importantly, we show that for specific phase patterns these non-Hermitian solutions present a spatial profile similar to that of topological edge states. The effect of such a phase-engineered drive is thus analogous to optically breaking the chain and creating a topologically non-trivial interface.

\noindent\textbf{The nonlinear driven-dissipative SSH model}

\noindent In presence of drive and dissipation, the physics of the nonlinear SSH model can be captured by a discretized Gross-Pitaevskii equation:
\begin{equation}
	    i\hbar\frac{d}{dt}\begin{bmatrix}a_{n}\\ b_{n}\end{bmatrix} =
	    \left( E_{0} -i\frac{\gamma}{2}\right)\begin{bmatrix}a_{n}\\ b_{n}\end{bmatrix}
	    +g\begin{bmatrix}|a_{n}|^{2}a_{n}\\ |b_{n}|^{2}b_{n}\end{bmatrix}\nonumber
	    -J\begin{bmatrix}b_{n}\\ a_{n}\end{bmatrix}
	    -J'\begin{bmatrix}b_{n-1}\\ a_{n+1}\end{bmatrix}
	    +i\begin{bmatrix}F_{a,n}\\F_{b,n}\end{bmatrix}e^{iwt}
    \tag{1}
    \label{eq2}
\end{equation}
\noindent where ${\bm \psi}_{n}=\left[a_{n};b_{n}\right]^{T}$ is a spinor describing the wavefunction of the A and B sites in the $n^{th}$ unit cell; $E_{0}$ is the on-site energy, $\gamma$ the decay rate, $g$ the interaction energy, $\hbar\omega$ the driving field energy, and $J$ ($J'$) is the intracell (intercell) coupling energy. The driving field ${\bm F}_{n}=\left[F_{a,n};F_{b,n}\right]^{T}$ can be engineered with a specific amplitude and phase on each site, $F_{\alpha,n}=|F_{\alpha,n}|e^{i\varphi_{\alpha,n}}$.

The topological nature of the SSH model is related to chiral (or sub-lattice) symmetry\cite{Chiu2016}, which imposes identical on-site energies and restricts the coupling terms to sites belonging to distinct sub-lattices, e.g. no next-nearest neighbour coupling.  Zero-energy edge states which emerge in lattices ending with weak links are protected by this symmetry. They are localised on a single sub-lattice, thus presenting a sub-lattice pseudo-spin:
\begin{equation}
\tilde{S}=\frac{\sum_{n}|a_{n}|^2-|b_{n}|^2}{\sum_{n}|a_{n}|^2+|b_{n}|^2}=\pm1,
\tag{2}
\label{Eq:spin}
\end{equation}
\noindent The sign of the spin reflects the sub-lattice localization, either on $A$ (+1) or $B$ (-1). Hereafter, we will show that the pseudospin properties of gap solitons generated in the non-linear driven-dissipative SSH model are strongly linked to their robustness properties against defects.

To emulate this system with cavity polaritons, we design an array of coupled micro-pillars ($\mathrm{3~\mu m}$ diameter), with alternating short ($\mathrm{2.2~\mu m}$) and long ($\mathrm{2.75~\mu m}$) center-to-center distances, see Fig.~\ref{fig1} (a). This array is fabricated by etching an epitaxially grown semiconductor heterostructure that consists in a planar cavity embedding a quantum well (see Methods for more details). A schematic representation of a single micro-pillar is presented in Fig.~\ref{fig1}~(b), together with a representation of the first (second) lower energy mode profile with s-like (p-like) symmetry.
	
The linear spectrum of this structure can be probed by low temperature (4~K) photoluminescence experiments. Imaging the emission with angular or spatial resolution enables observing polaritonic bands either in momentum or real space, see Fig.~\ref{fig1} (c) and (d). The two bands are formed from the hybridization of the s mode of all pillars and emulate the single-particle SSH model. The asymmetry of the spectrum with respect to the center of the topological gap is attributed to couplings between s and p modes~\cite{Mangussi2020}.
	
\noindent\textbf{Formation of gap solitons}
	
\begin{figure}
\begin{center}
	\includegraphics[width=0.7\linewidth]{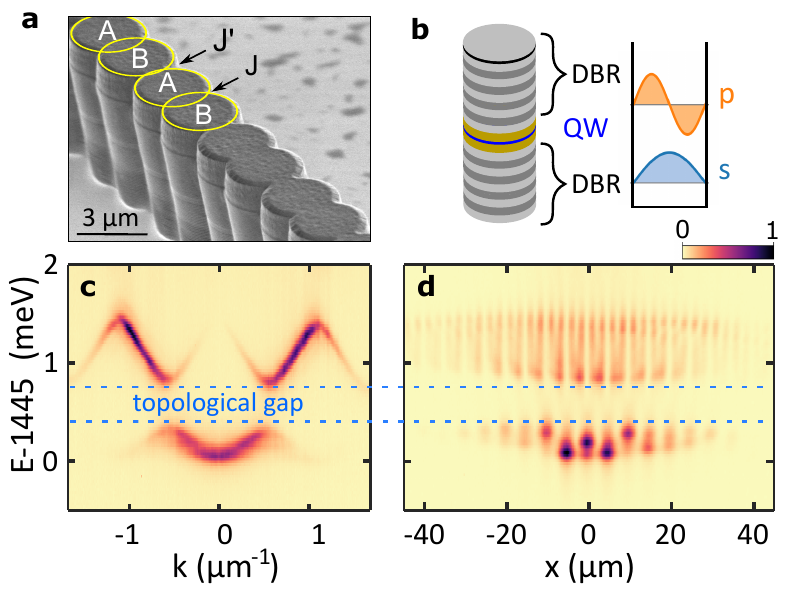}
	\end{center}
	\caption { \textbf{Implementation of the SSH lattice.}
			\textbf{a,}
	Scanning electron microscope image of the SSH polariton lattice. We highlight some of the micro-pillars with yellow circles.
			\textbf{b,}
	Left: Schematic representation of a micro-pillar, with a cavity (orange layer) containing a single quantum well (QW) embedded between two distributed Bragg reflectors (DBR). Right: Typical shape of the $s$ and $p$ polariton modes.
			\textbf{c-d,}
	Spectrally resolved photoluminescence intensity measured as a function of momentum $k$ (\textbf{c}) and position $x$ along the lattice (\textbf{d}).
		}
	\label{fig1}
\end{figure}
	
\noindent To probe the polariton nonlinear response, we implement a quasi-resonant excitation spectrally tuned to the center of the topological gap, and we measure the transmitted intensity. The excitation spot ($3~\mu$m FWHM) is focused at the center of a dimer located more than ten unit cells away from the lattice edges. This excitation scheme corresponds, in Eq.~(1), to a driving field localized on a single dimer with equal amplitude and phase on both A and B sites.
	
\begin{figure}
\begin{center}
    \includegraphics[width=0.7\linewidth]{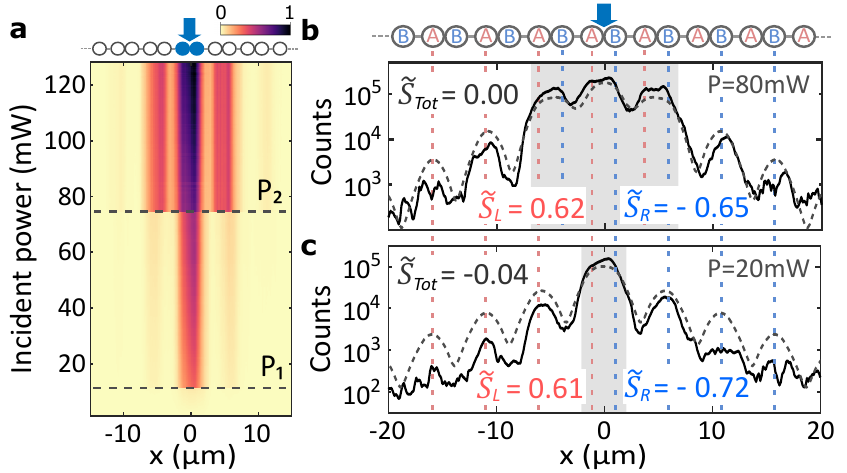}
    \end{center}
		\caption { \textbf{Generation of topological gap solitons.}
			\textbf{a,} Spatially resolved intensity profiles, integrated along the direction perpendicular to the lattice, measured as the driving power is ramped up. Intensity is normalized to its maximum value. Dashed horizontal lines mark the $P_1$ and $P_2$ expansion threshold. The driven dimer is schematically shown on top. \textbf{b-c,} Experimental (solid line) and simulated (dashed line) spatial profiles of the soliton above $P_2$ (\textbf{b}) and $P_1$ (\textbf{c}) thresholds. For each dimer, red and blue dashed lines mark the pillar presenting a strong emission. Gray areas indicate high-intensity regions.}
		\label{fig2}
\end{figure}
	
As we ramp up the driving power, we observe in real space images the formation of high intensity domains that are symmetric with respect to the pump and localized on a discrete number of dimers (see Fig.~\ref{fig2}~(a)). When the incident power reaches a first threshold denoted $P_{1}$, we observe the formation of a one-dimer bright domain. Above threshold $P_{2}$, the high-intensity domain abruptly extends to three dimers.
	
Each threshold occurs when polaritons locally enter the nonlinear regime, i.e. when the interaction energy within a dimer overcomes the spectral detuning between the pump and the top of the lower band. These nonlinear solutions correspond to gap solitons, which have been observed in various platforms including atomic\cite{Eiermann2004}, photonic\cite{Sukhorukov2002, Kanshu2012} and polaritonic\cite{Tanese2013, Cerda-Mendez2013, Goblot2019} lattices. They are composed of a high-intensity region (grey-shaded areas in  Fig.~\ref{fig2}) and of exponentially decaying tails on both sides (white areas). As a result of the lattice inversion symmetry, the sub-lattice pseudospin computed over the entire profile vanishes: $\tilde{S}_{\mathrm{tot}}= -0.04\pm 0.05$ after $P_{1}$ and $\tilde{S}_{\mathrm{tot}}= 0.00 \pm 0.02$ after $P_{2}$.
	
It is insightful to probe how the pseudospin locally varies over the profile. Inside the high-intensity domains, each dimer shows similar amplitude on both sub-lattices, leading to a locally vanishing pseudospin~\cite{Smirnova2019} (see Fig.~\ref{fig2}~(b-c)). This is a direct consequence of the fact that these domains are formed by bonding modes of a discrete number of dimers, blueshifted by the polariton-polariton interaction. In contrast, the evanescent tails are strongly localized on a single sub-lattice: the $A$ sub-lattice on the left, and the $B$ on the right. Consequently, the pseudo-spins $\tilde{S}_L$ and $\tilde{S}_R$ integrated over the left and right soliton tails do not vanish and present opposite signs: $\tilde{S}_{L}>0$ and $\tilde{S}_{R}<0$ (see Fig.~\ref{fig2}~(b-c)). This feature can be understood by considering that the high density region locally breaks the lattice symmetry and acts as a defect. Moreover this defect is connected to the lattice \textit{via} a weak link on both sides. As a result, the soliton evanescent tails are similar to that of topological edge states with characteristic pseudo-spin polarization.
	
To accurately reproduce experimental data, we use an effective 1D continuous model, which allows taking into account the finite size of the lattice pillars and the mixing of s-p bands. We look for steady-state solutions of the following Gross-Pitaevskii equation:
\begin{equation}
    i\hbar\frac{d\psi(x,t)}{dt}=\left[-\frac{\hbar^2}{2m} \nabla^2 + V(x) \right]\psi(x,t)+g|\psi(x,t)|^2 \psi(x,t)\nonumber
    -i\frac{\hbar\gamma}{2}\psi(x,t)+i F(x) e^{-i \omega_{p} t}
    \tag{3}
    \label{tSch}
\end{equation}
where $m$ is the polariton mass. The strong and weak links are respectively represented by barriers of small and large amplitudes in the potential $V(x)$ (see Supplementary section~1). Numerical results shown in Fig.~\ref{fig2} reproduce the measured soliton profiles with their characteristic tails of opposite spin polarization. Note that the chiral symmetry of the polariton lattice is perturbed by the s-p coupling, explaining the asymmetry between the two bands in Fig.~\ref{fig1}~(c). Nevertheless, the effective coupling between pillars being an order of magnitude smaller than the energy splitting between s and p orbitals, deviations from the perfect chiral system do not significantly alter sublattice spin properties of topological solitons investigated in the present work.

\begin{figure}
    \begin{center}
	\includegraphics[width=0.44\linewidth]{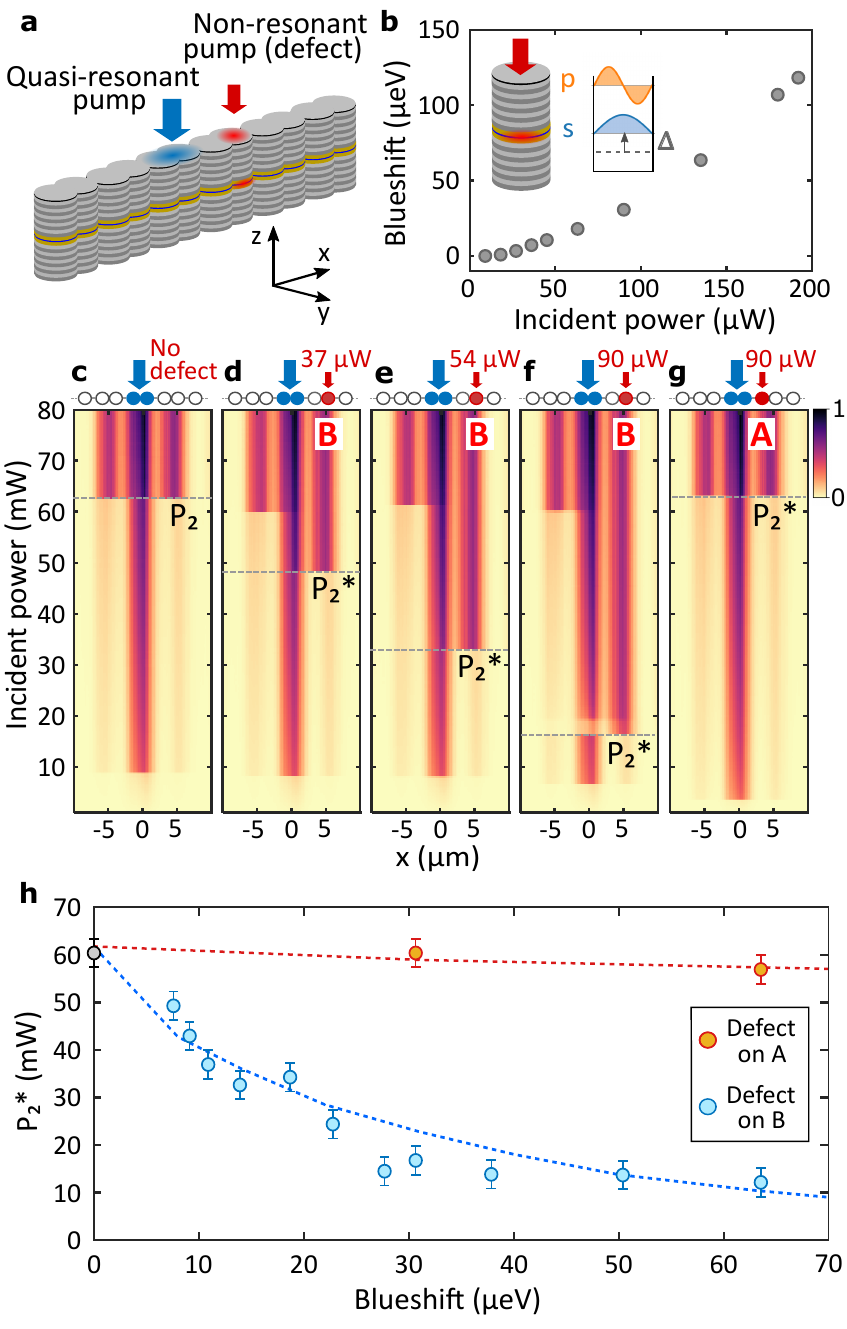}
    \end{center}
	\caption { \textbf{Probing the robustness of topological solitons against a defect.}
		\textbf{a,}
		Scheme of the two beam experiment to generate both a soliton (quasi-resonant pump, blue arrow) and an optically induced defect (off-resonant pump, red arrow). \textbf{b,} Calibration of the optical defect: blueshift $\Delta$ as a function of non-resonant pump power measured on an isolated micro-pillar.
		\textbf{c-g,}
    	Integrated intensity spatial profiles measured as a function of the resonant driving power for (\textbf{c}) no defect, (\textbf{d-f}) a defect on the $B$ sub-lattice generated with various non-resonant powers, and (\textbf{g}) a defect on the $A$ sub-lattice generated with same power as in (\textbf{f}). For each panel, the driven dimer (defect) is shown on top with blue (red) circles. \textbf{h,} Threshold power $P_2^*$ measured as a function of the blueshift (deduced from \textbf{b,}) induced by a defect localized on the $A$ (red symbols) or $B$ (blue symbols) sub-lattice. Red (blue) line corresponds to numerical simulations.
		}
	\label{fig3}
\end{figure}
	
\noindent\textbf{Robustness of gap solitons in the presence of a non-Hermitian defect}
	
\noindent It was theoretically predicted for conservative SSH lattices\cite{Solnyshkov2017}, that these spin-polarized tails play a crucial role in the robustness properties of topological solitons, and their interaction with a local defect. Hereafter, we extend this idea to the non-Hermitian context, and show that the nonlinear thresholds for the lateral growth of the solitons are drastically affected by the presence of a defect on one sub-lattice but not on the other one.
	
In order to probe the effect of a perturbation, we consider a defect on a pillar belonging to a neighboring dimer of the soliton. We then monitor how the spatial expansion of the soliton is affected by this perturbation. The defect is optically generated by non-resonantly pumping a pillar belonging to the A or B sublattice with a second laser (see Fig.~\ref{fig3} a). This creates a local reservoir of excitons whose effect is twofold: 1- it locally blueshifts the pillar on-site energy thus acting as a perturbation on the real part of the potential; and 2- it induces a local gain through stimulated relaxation of excitons toward polariton modes, thus acting as an imaginary perturbation of the potential. This defect is therefore intrinsically non-Hermitian.
	
To calibrate the real part of this perturbation, we measure the spectral shift of the ground-state emission from a single micro-pillar presenting the same characteristics as the ones forming the chain (see Fig.~\ref{fig3}~(b)). For the imaginary part, we evaluate the lasing threshold in the same pillar ($P_{\rm th}\sim0.6~\mathrm{mW}$), which enables us to estimate the gain induced by the non-resonant pump creating the defect (see Supplementary section~3).
	
The evolution of the spatial profile of solitons as we ramp up the resonant pump power is displayed in Fig.~\ref{fig3} (d)-(g) for different amplitudes and positions of the defect. For comparison, the case with no defect amplitude is shown in Panel~(c). We first monitor the soliton expansion in presence of a defect located on a B sub-lattice site, where the soliton intensity presents a local maximum. The experimental profiles are shown in Panels (d)-(f) (see Supplementary section~3 for the calculated profiles). As the defect breaks the system spatial symmetry with respect to the excitation spot, and locally reduces the laser detuning, the soliton expansion becomes asymmetric. It is first favored toward the defect at power $P_{2}^{*}$. At a higher power, close to power $P_{2}$ measured without defect, the soliton eventually recovers a symmetric profile. We observe in Fig.~3~(h) that $P_{2}^{*}$ strongly varies with the defect amplitude. For instance, a defect as small as a third of the polariton linewidth ($20~\mu\mathrm{eV}$) is sufficient to reduce $P_{2}^{*}$ by a factor of two with respect to $P_{2}$. In the simulated curve shown in Fig.~3~(h), this strong sensitivity is well accounted for by including both calibrated values of the real and imaginary parts of the perturbation (see Supplementary section~3).

This high sensitivity to defects located on the $B$ sub-lattice contrasts with what we measure when the defect is localized on the $A$ sub-lattice, where the amplitude of the tails vanishes. In that case, the second nonlinear expansion simultaneously occurs toward the right and left dimer, regardless of the defect amplitude (see Fig.~\ref{fig3}~(g)-(h)) and for a power close to the one measured in absence of defect. This difference in sensitivity to defects located on A or B sublattice is a direct consequence of the spin polarization of the soliton tails. These results demonstrate that the lattice chiral symmetry provides topological gap solitons with high robustness against non-Hermitian defects located on one sub-lattice.
	
\begin{figure}
\begin{center}
	\includegraphics[width=0.7\linewidth]{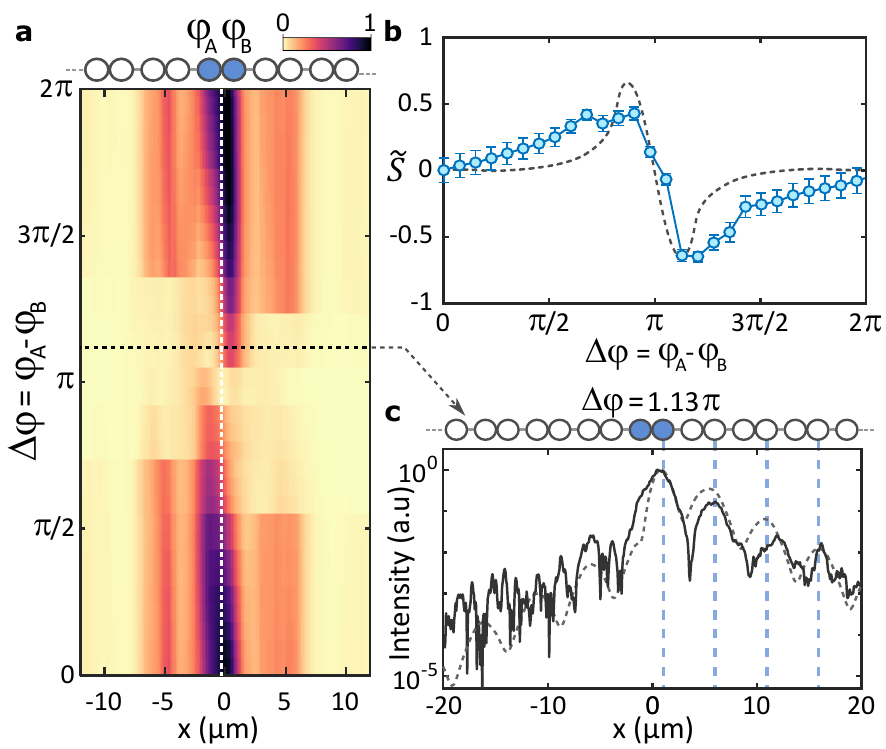}
	\end{center}
	\caption { \textbf{Spin-polarized topological solitons.}
		\textbf{a,} Integrated spatial profile measured as a function of $\delta \varphi$ for a total excitation power $P > P_2$. The driven dimer are depicted on top (blue dots), and the white dashed line indicates the center of the driven dimer. Horizontal dashed line indicates the profile shown in (\textbf{c}). \textbf{b,} Measured (symbols) and calculated (dashed line) soliton global pseudospin as a function of $\Delta \varphi$ \textbf{c,} Intensity profile of the soliton measured (solid line) and simulated (dashed line) for a phase difference $\Delta\varphi = 1.13~\pi$ (indicated by the dashed line in Panel \textbf{a}). Blue dashed lines mark B pillars with high intensity.}
	\label{fig4}
\end{figure}
	
\noindent\textbf{Controlling the pseudospin of solitons with a phase-engineered drive}
	
\noindent The solitons we have considered so far present a globally vanishing pseudospin, like in conservative systems\cite{Smirnova2019}. Hereafter, we show how we can depart from this family of unpolarized solutions thanks to the driven-dissipative nature of polaritons. To do this, we turn to an excitation scheme where two pillars of a dimer are driven with two beams with same amplitude and tunable phase difference $\Delta \varphi=\varphi_{\mathrm{A}}-\varphi_{\mathrm{B}}$ (see inset above Fig.~\ref{fig4}~(a)). In Eq.~(1), this corresponds to a driving field still localized on a single dimer with equal amplitude on both sub-lattices, but different phases. We select a total excitation power $P>P_2$ in order to obtain a three-dimer soliton when $\Delta \varphi=0$. Fig.~\ref{fig4}~(a) presents the evolution of the intensity distribution along the lattice upon increasing $\Delta \varphi$. A complex sequence of abrupt switchings between distinct regimes with different numbers of bright pillars is revealed. Starting from $\Delta \varphi = 0$, we observe successive transitions from three bright dimers all the way to the linear regime as we approach $\Delta \varphi = \pi$.

Interestingly, for non-zero values of $\Delta \varphi$, the nonlinear intensity patterns exhibit imbalance between the two sublattices. Such imbalance is due to the phase gradient imposed by the driving field which breaks the inversion symmetry of the system. The measured and calculated evolutions of the global soliton pseudospin are presented in Fig.~\ref{fig4}~(b). Both exhibit a clear evolution toward $1$ ($-1$) as the phase difference approaches $0.90~\pi$ ($1.13~\pi$). This entire set of spin-polarized solutions does not exist in conservative systems, as they require the phase-patterned driving field to be stabilized (see Supplementary section~4). They can thus be considered as intrinsically non-Hermitian.

As the phase difference approaches $\pi$, the driving field becomes orthogonal to the Bloch modes in the lowest band thus strongly reducing light injection in the lattice. Remarkably, just before its complete extinction, the high-intensity part of the solitons is restricted to a single pillar of the driven dimer, while the other one experiences destructive interference. We show experimental and theoretical spatial profiles of the soliton for $\Delta\varphi = 1.13~\pi$ in Fig.~\ref{fig4}~(c). Interestingly the profiles present a significant evanescent tail on one side only, while pillars on the other side show vanishing intensity.
This can be understood by the fact that the destructive interference effectively tends to decouple the bright soliton from one half of the lattice. Consequently, the soliton is now connected to the lattice \textit{via} a weak link on one side only (thus showing a spin polarized exponential tail on that side), while the coupling to the other side of the lattice is strongly reduced. Hence the entire profile of the gap soliton is alike that of a topological edge state.

To the best of our knowledge, these spin-polarized states whose existence requires the combined interplay of nonlinearity, non-Hermiticity and topology have never been considered so far. Their emergence can be understood by the onset of a phase frustration between the pump and the polariton fluid in the driven dimer, leading to vanishing amplitude in one of the two pillars. As a result, the effect of this frustration is similar to an optical suppression of one half of the lattice, and generation of a topologically non-trivial interface~\cite{Smirnova2019}.

\noindent\textbf{Outlook}
	
\noindent This work provides new perspectives to nonlinear topological photonics. Indeed, we have shown that the non-hermitian nature of photonic systems provides a general way to optically generate topological interfaces in the bulk of lattices, to stabilize nonlinear solutions that find no equivalent in conservative systems, or to generate non-Hermitian defects. It will be of the highest interest to explore nonlinear topological physics in more complex non-Hermitian systems with chiral symmetry, like driven-dissipative Lieb \cite{Klembt2017, Whittaker2018, Goblot2019} and honeycomb \cite{Real2020, Suchomel2018} lattices or higher-order topological insulators\cite{Mittal2019, ElHassan2019}. Phase-engineered drives may offer the opportunity to create non-trivial interfaces in the bulk of these lattices, and generate nonlinear excitations that have no counterparts in conservative topological systems. Such approach will be useful for future exploration of many-body topological effects in open systems\cite{Diehl2011,Bardyn2013, Konotop2016, Fraser2019}.

\begin{methods}

\noindent \textbf{Sample description}

\noindent The periodic structure used in this work is etched out of a planar semiconductor microcavity with high quality factor ($Q \approx 75,000$) grown by molecular beam epitaxy. The microcavity is composed of a $\lambda$ GaAs layer embedded between two $\mathrm{Ga_{0.9}Al_{0.1}As/Ga_{0.05}Al_{0.95}As}$ distributed Bragg reflectors with 32 (top) and 36 (bottom) pairs. A single $\SI{8}{\nano \meter}$ $\mathrm{In_{0.05}Ga_{0.95}As}$ quantum well is inserted at the center of the cavity, resulting in strong exciton-photon coupling, with an associated $\SI{3.5}{\milli \eV}$ Rabi splitting. After epitaxy, the sample is processed with electron beam lithography and dry etching into arrays of concatenated pillars arranged in a SSH lattice. The exciton-photon detuning, defined as energy difference between the uncoupled planar cavity mode and the exciton resonance, is of the order of $\Delta_{C-X} \approx \SI{6.5}{\milli \eV}$ for all the experiments.
	
\noindent \textbf{Experimental techniques}

\noindent The sample is cooled down to $T = \SI{4}{K}$. Non-resonant photoluminescence measurements are realized with a single-mode continuous-wave laser at 780 nm. The excitation spot is elongated (${\rm FWHM}\sim30~\mu\mathrm{m}$) using a cylindrical lens. The emission is collected through a lens with NA 0.65 and imaged on the entrance slit of a spectrometer coupled to a charge-coupled device camera with $\sim 30~{\mu \rm{eV}}$ spectral resolution. Real- and momentum-space photoluminescence images are realized by imaging the sample surface and the Fourier plane of the objective, respectively. A polarizer is used to select emission polarized along the long axis of the lattice. Experiments with quasi-resonant excitation are realized in transmission geometry, with the excitation (detection) on the epitaxial (substrate) side of the sample and a spot of $3.5~ \mu$m FWHM. The optical defect is created by focusing onto a $3~ \mu$m FWHM spot a $\SI{825}{\nano \meter}$ cw laser on the epitaxial side. For the two-spot experiment (each with $3.5~ \mu$m FWHM), the phase difference is induced with a delay line with one of the mirrors mounted on a piezoelectric actuator. Error bars in Fig.~3~(h) correspond to the standard deviation on the measurement of $P_2$ obtained by performing a repeatability study on the resonant pump alignment. Error bars in Fig.~4~(b) are calculated by evaluating the impact of modifications in the area chosen for intensity integration on the pseudo spin calculation.

\noindent \textbf{Numerical methods}

\noindent We use the 3-rd order Adams-Bashforth method for numerical integration of the nonlinear Gross-Pitaevskii Eq.~\eqref{tSch} with a time step of $10^{-3}$~ps. CPU-based parallel computing is used to evaluate the kinetic energy term via Fast Fourier Transform. The numerical grid of $2^9=512$ points allows to describe a lattice of 21 dimers with a step of $0.25$~$\mu$m. Increasing the resolution further or considering a longer chain does not change the results. The ramp-up and ramp-down times are 20~ns, which is a compromise between the required adiabaticity and simulation time. We have checked that increasing the ramp-up time further does not change the results. The parameters were as follows: $m=3\times 10^{-5}m_0$ ($m_0$ is the free electron mass), $\hbar\gamma=70$~$\mu$eV, $g=5~ \mu {\rm eV} \cdot \mu {\rm m}^2$, pumping spot FWHM $\approx 2.8$~$\mu$m, and the pumping frequency detuning with respect to the band edge is $\approx 0.2$~meV. Other relevant parameters, such as the potential profile, are provided and discussed in the Supplementary materials.
	
\end{methods}

\bibliographystyle{naturemag}

\begin{addendum}
\item This work was supported by the Paris Ile-de-France R\'egion in the framework of DIM SIRTEQ, the Marie Sklodowska-Curie individual fellowship ToPol, the EU project "QUANTOPOL" (846353), the H2020-FETFLAG project PhoQus (820392), the QUANTERA project Interpol (ANR-QUAN-0003-05), the French National Research Agency project Quantum Fluids of Light (ANR-16-CE30-0021),the French RENATECH network, the French government through the Programme Investissement d’Avenir (I-SITE ULNE / ANR-16-IDEX-0004 ULNE) and IDEX-ISITE initiative 16-IDEX-0001 (CAP 20-25), managed by the Agence Nationale de la Recherche, the Labex CEMPI (ANR-11-LABX-0007), the CPER Photonics for Society P4S and the M\'etropole Europ\'eenne de Lille (MEL) via the project TFlight.
\item[Author contributions]
N.P. and P.S.-J. performed the experiments and analyzed the data. N.P. performed initial theoretical modelling of the experiments using the tight-binding approach, which led to the discovery of spin-polarized topological solitons. D.D.S and G.M provided theoretical guidance and performed the theoretical calculations in the 1D continuous model. N.P., P.S.-J., D.D.S., G.M, N.C.Z., B.R., O.J. A.A., S.R. and J.B. participated to scientific discussions. N.P., P.S.-J., D.D.S., G.M, A.A., S.R. and J.B. wrote the manuscript. N.C.Z. and B.R. contributed to editing of the manuscript. P.S.-J., S. R., J. B. and A.A. designed the sample. A.L., L.L.G., T.B., A.H. and I.S. fabricated the samples. A.A, S.R. and J.B. supervised the work.
\item[Competing Interests] The authors declare that they have no
competing financial interests.
\item[Correspondence] Correspondence
should be addressed to jacqueline.bloch@c2n.upsaclay.fr
\end{addendum}

\newpage
\setcounter{figure}{0}
\setcounter{equation}{0}
\renewcommand{\thefigure}{S\arabic{figure}}
\renewcommand{\theequation}{S\arabic{equation}}
\renewcommand{\thefootnote}{\fnsymbol{footnote}}
\begin{center}
\title{\begin{large}Supplementary materials for the paper entitled: Topological gap solitons in a 1D non-Hermitian lattice\end{large}}
\end{center}
\author{N.~Pernet$^{1}$, P.~St-Jean$^{1}$\footnote[1]{These authors contributed equally: N.~Pernet, P.~St-Jean}, D.D.~Solnyshkov$^{2,3}$, G.~Malpuech$^2$, N. Carlon Zambon$^1$, B.~Real$^4$, O.~Jamadi$^4$, A.~Lema\^itre$^1$, M. Morassi$^1$, L.~Le~Gratiet$^1$, T.~Baptiste$^1$, A.~Harouri$^1$, I.~Sagnes$^1$, A.~Amo$^4$, S.~Ravets$^1$, J.~Bloch$^1$}


\maketitle
	
\begin{affiliations}
\item Universit\'{e} Paris-Saclay, CNRS, Centre de Nanosciences et de Nanotechnologies (C2N), 91120, Palaiseau, France
\item Institut Pascal, PHOTON-N2, Universit\'e Clermont Auvergne, CNRS, SIGMA Clermont, F-63000 Clermont-Ferrand, France.
\item Institut Universitaire de France (IUF), 75231 Paris, France
\item Universit\'{e} de Lille, CNRS, Laboratoire de Physique des Lasers Atomes et Mol\'{e}cules (PhLAM), 59000 Lille, France
\end{affiliations}


\section{The 1D effective model mapping the SSH polariton lattice}

\subsection{\textit{The 1D potential.}}

While the exact description of the chain of polariton micropillars would require solving 3D Maxwell's equations, such solution would be numerically too demanding and time consuming. Actually, the fastest and simplest description of the system would be given by the tight-binding approximation described by Eq.~(1) of the main text. Here, we use a less approximate description based on an effective 1D potential. The quantization of the polariton modes in the vertical direction makes them massive and allows using a 1D effective Schrödinger equation considering only the propagating direction along the lattice. This allows taking into account band-mixing effects and the finite size of the lattice sites. The 1D landscape for the effective potential is constituted by a series of potential wells separated by two types of energy barriers. The strong and weak links are respectively represented by barriers of small and large amplitudes (corresponding to strongly and weakly overlapping pillars), as shown in Fig.~\ref{figS1}~(a). The corresponding Hamiltonian therefore reads:
\begin{equation} \label{nf_hamiltonian}
		\mathcal{H}^{\mathrm{(1D)}}_{0}(x)=\left[ - \frac{\hbar^2}{2m} \nabla^2 + V(x) \right]
\end{equation}
where $m$ is the polariton mass and $V(x)$ is the aforementioned energy potential.

\subsection{\textit{Simulation of the polariton dispersion shown in Fig.1 of the main text.}}

\begin{figure}
\begin{center}
\includegraphics[width=0.7\columnwidth]{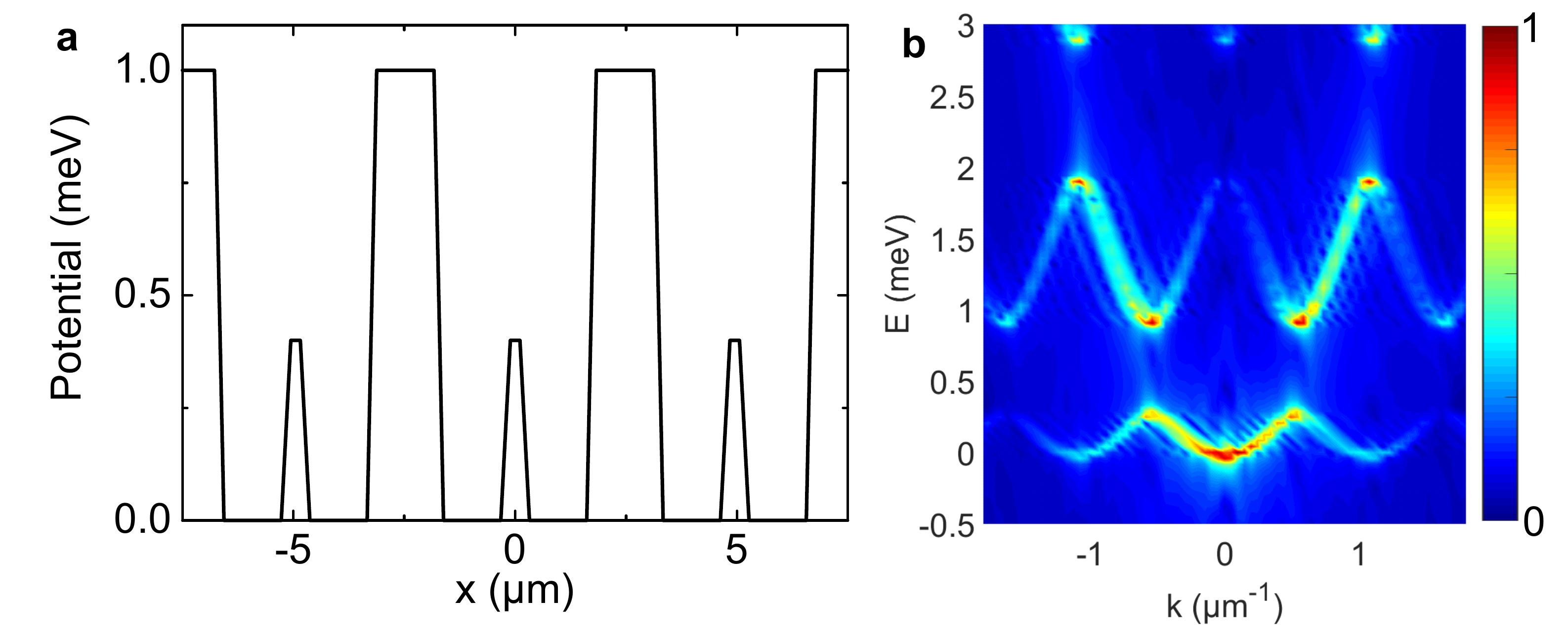}
\end{center}
\caption {a) Effective 1D potential describing an SSH chain; b) Dispersion obtained in numerical simulations.
}
\label{figS1}
\end{figure}

For convenience, we remind here the time-dependent 1D Gross-Pitaevskii equation (Eq.~(3) of the main text) that we use to simulate the experimental results:
\begin{eqnarray}
    i\hbar\frac{d\psi(x,t)}{dt}=\mathcal{H}^{\mathrm{(1D)}}_{0}(x)\psi(x,t)+g|\psi(x,t)|^2 \psi(x,t)
    -i\frac{\hbar\gamma}{2}\psi(x,t)+i F(x) e^{-i \omega_{p} t}
    \label{tSch}
\end{eqnarray}
The simulated band structure of the polariton modes in the 1D potential $V(x)$ is obtained by solving this equation without interactions ($g=0$) for a localized excitation at $t=0$: $\psi(x,0)=\exp(-x^2/2\sigma^2)$, where the width of the pulse has to be sufficiently small $\sigma=0.5$~$\mu$m. It is represented in Fig.~\ref{figS1}~(b) and shows a qualitative agreement with the one obtained experimentally. The size of the gap and the effective mass of the lower band are well reproduced. The differences between the experimental and the simulated bands visible at higher energies are most probably due to the 2D nature of the actual system.

\section{Simulation of gap soliton spatial profiles shown in Fig.2 of the main text}

\begin{figure}
\begin{center}
\includegraphics[width=0.3\columnwidth]{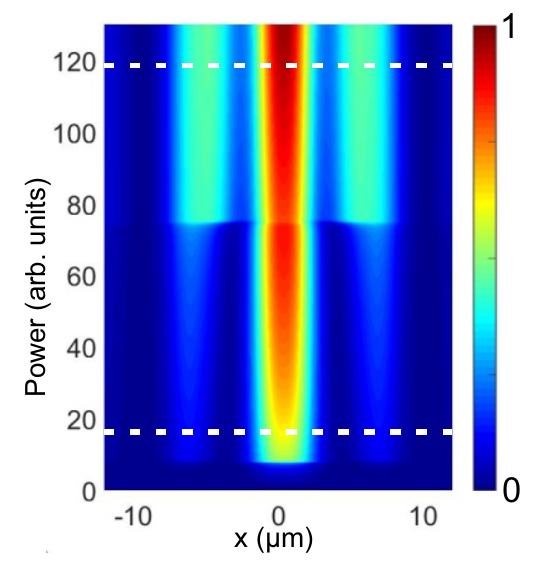}
\end{center}
\caption {Numerical simulations showing the spatial distribution of the polariton density $|\psi(x)|^2$ at different pumping powers when scanning up the power. Dashed horizontal lines indicate the simulated profiles shown in Fig.~2~(b-c) of the main text.}
\label{figS2}
\end{figure}

Solving Eq.~(S2) with the parameters given in Numerical Methods of the main text and with a linear increase of the pumping power determined by $|F|^2$, we obtain the data shown in Fig.~\ref{figS2}, which nicely reproduce the experimental observations. In particular, we observe the formation of a 1D dimer soliton above a first threshold, and then above a second threshold the formation of a symmetric three dimer soliton. The simulated profiles shown in Fig.~2~(b-c) of the main text corresponds to powers indicated with dashed horizontal lines.

\section{Numerical study of the soliton robustness against a defect induced by an off-resonant pump}

Here, we describe the results of numerical simulations, based on Eq.~\eqref{tSch}, corresponding to the experimental measurements shown in Fig.~3~(c-g) of the main text. To take into account the optically induced defect, we add to $V(x)$ an additional potential $V_D(x)=(\Delta + i \Gamma) G(x)$, where $G(x)$ is a Gaussian with 3~$\mu$m FWHM representing the spot shape. $\Delta$ is the defect amplitude measured experimentally on a reference micropillar. $\Gamma$ is a positive imaginary part describing stimulated scattering from the reservoir into the polariton mode. The amplitude of this imaginary part is calibrated considering the ratio of the pumping power to the threshold power for polariton lasing in the reference micropillar (30\% of the lasing threshold power for a 100~$\mu$eV blueshift).

\begin{figure}
\begin{center}
\includegraphics[width=0.75\columnwidth]{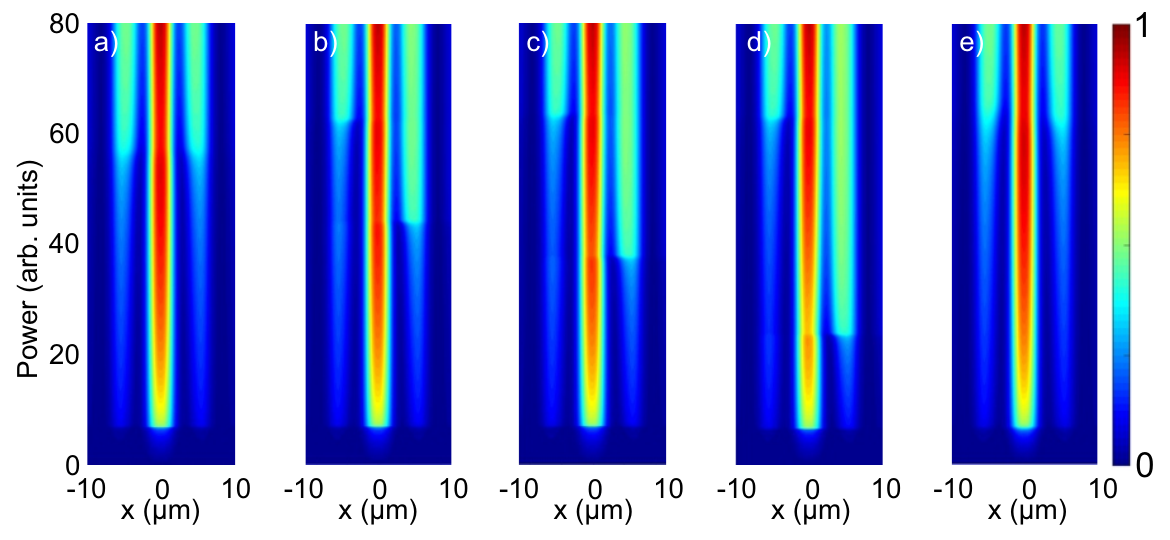}
\end{center}
\caption {Numerical simulations showing the spatial distribution of the polariton density $|\psi(x)|^2$ at different pumping powers, for the following defect potentials ($\mu$eV) a) 0 (B-site), b) 8 (B-site), c) 12 (B-site), d) 20 (B-site), e) 20 (A-site).
}
\label{figS3}
\end{figure}

\begin{figure}
\begin{center}
\includegraphics[width=0.5\columnwidth]{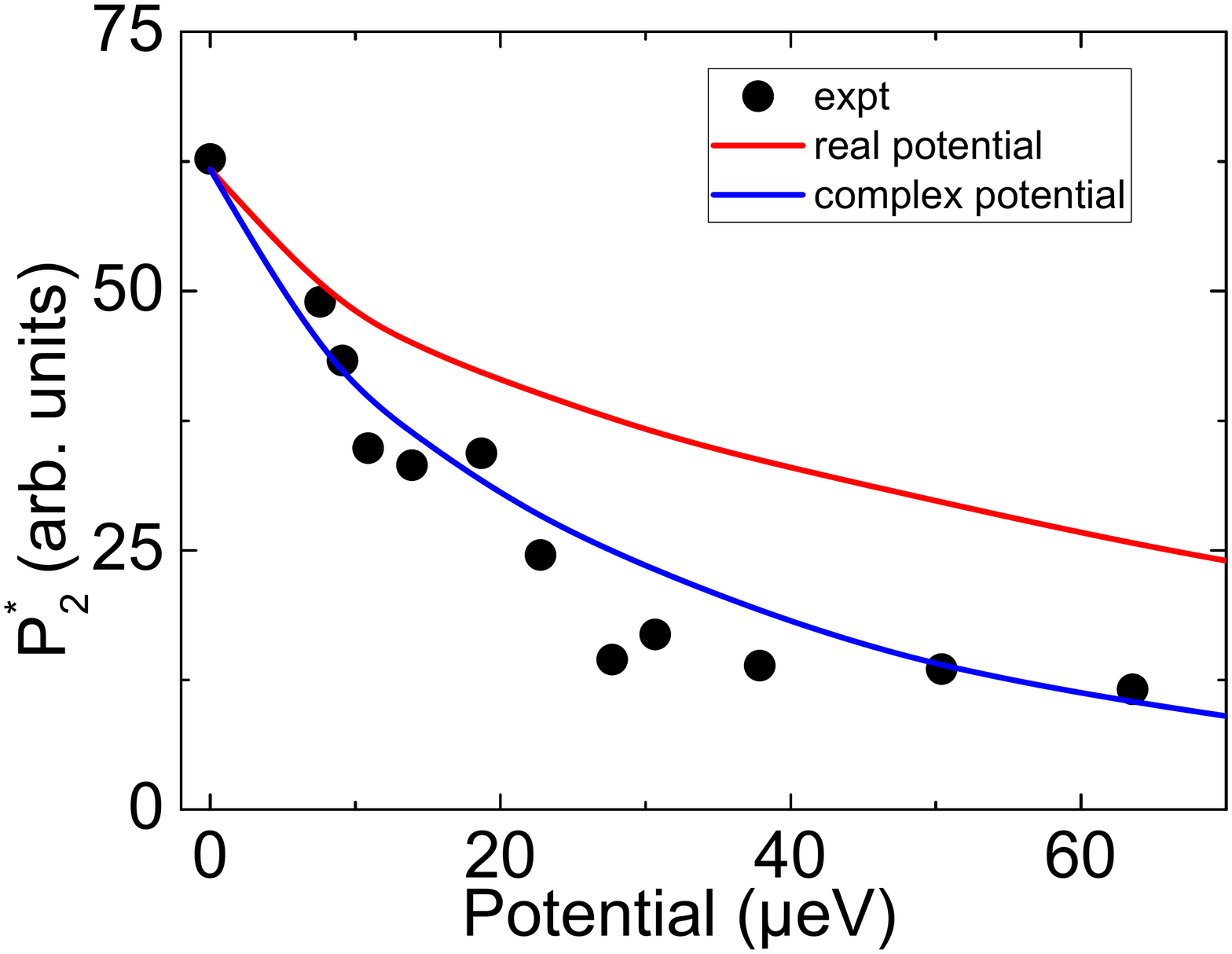}
\end{center}
\caption {Values of the threshold power $P_2^*$ obtained with a defect located on a B site in the experiment (black dots), simulated taking into account the real part of the defect potential only (red line) or taking into account both its real and imaginary parts (blue line).
}
\label{figS3RC}
\end{figure}

The simulated intensity profiles are shown in Fig.~\ref{figS3}~(a-e) for the following defect amplitudes (expressed in $\mu$eV): 0 (a), 8 (b), 12 (c), 20 on the B-site with high soliton density (d) and 20 on the A-site with low soliton density (e). The calculated profiles show good agreement with the measured ones. To quantitatively compare theory with experiments we report in Fig.~\ref{figS3RC} the measured values of $P_2^*$ and the simulated values taking into account only the real part of the defect or both real and imaginary parts. We see that the gain leads to a significant reduction of the threshold. The agreement with experiments is better when including both real and imaginary parts of the defect, that is when taking into account the non-Hermitian nature of the defect.

\section{Simulation of spin polarized solitons}

We discuss here simulation of experiments reported in the last part of the main text. The lattice is resonantly driven with two pumping spots focused on the two pillars composing the central dimer. The pumps have same amplitude but phase difference $\Delta \varphi$. The steady state non-linear intensity profiles strongly depend on $\Delta \varphi$.

\subsection{\textit{Phase scan and profiles.}}

\begin{figure}
\begin{center}
\includegraphics[width=0.25\columnwidth]{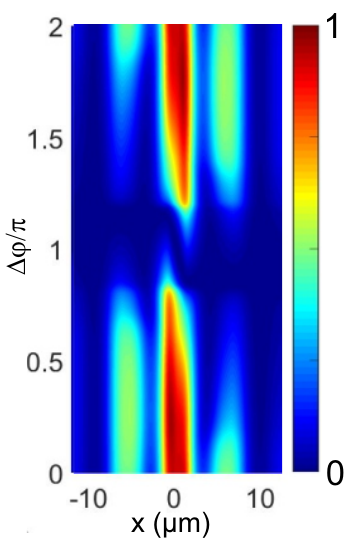}
\end{center}
\caption {Numerical simulations showing the spatial distribution of the polariton density $|\psi(x)|^2$ as a function of phase difference $\Delta \varphi$ between the two pumping lasers on A and B sites of the same unit cell.
}
\label{figS4}
\end{figure}

In this subsection, we show the results of numerical simulations corresponding to the experiment shown in Fig.~4~(a) of the main text. We vary the relative phase $\Delta\varphi$ between the two pumps between $0$ and $2\pi$, while keeping the pump intensity constant. Numerical results are presented in Fig.~\ref{figS4}. They reproduce the overall experimental behavior with a complex sequence of abrupt switchings between distinct regimes with different numbers of bright pillars (starting from $\Delta \varphi = 0$, we successively observe three bright dimers, two bright dimers, three bright pillars, and only one bright pillar before switching to the linear regime as we approach $\Delta \varphi = \pi$). Interestingly, the simulations reproduce the establishment of non-linear solutions with non-zero global spin polarization: In the vicinity of $\Delta \varphi = \pi$, we obtain a highly spin polarized soliton, with a single bright pillar and a profile similar to a topological edge state.

Note that for these simulations, we have used the pump power as a fitting parameter. We have selected a value that reproduces best the experimental results, approximately equivalent to $40$~mW, with three dimers in the high density regime at $\Delta\varphi=0$. Changing this intensity affects the phase threshold close to $\Delta \varphi = \pi$ and modify the level of asymmetry in the intensity spatial distribution.

\subsection{\textit{Stability of the spin polarized solitons in absence of drive and dissipation.}}

\begin{figure}
\begin{center}
\includegraphics[width=0.6\columnwidth]{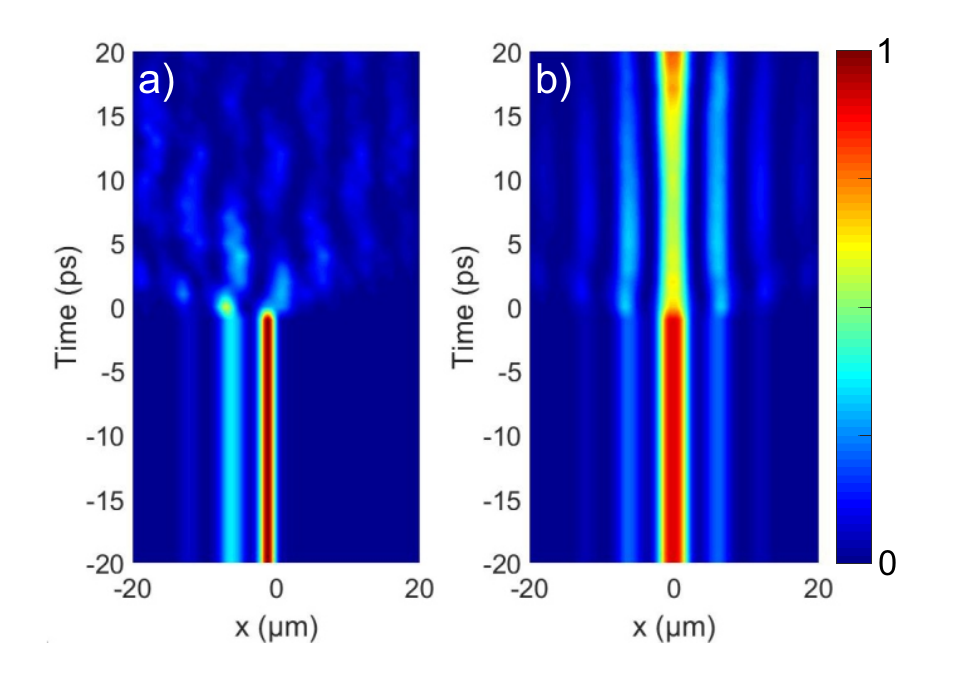}
\end{center}
\caption {Numerical simulations showing the spatial distribution of the polariton density $|\psi(x)|^2$ as a function of time. The pumping and decay terms are artificially turned off at $t=0$. a) asymmetric configuration with $\Delta\varphi=0.94\, \pi$ (corresponding to Fig.~4 of the main text) is unstable; b) symmetric configuration is close to the conservative gap soliton and remains relatively stable, only exhibiting oscillations.
}
\label{figS5}
\end{figure}

The strongly asymmetric solutions observed for a phase difference $\Delta\varphi\approx\pi$ can only be observed thanks to the driven-dissipative polariton configuration that we use. This is a very important difference with an ordinary topological gap soliton, which can be studied in a conservative system, such as optical waveguides. In this section, we illustrate numerically this property by monitoring the time evolution of the system after we abruptly switch off drive and dissipation in Eq.~\eqref{tSch}. As shown in Fig.~\ref{figS5}(a), the asymmetric spin polarized soliton, obtained for $\Delta\varphi$ close to $\pi$, is not a stationary solution any more. Moreover, being far from any stationary solution, it decays very rapidly. To provide a comparison, we also show a simulation for $\Delta\varphi=0$, in which case the solution is much closer to a conservative topological gap soliton (although the central part in the bistable regime has a higher density than that expected in the conservative case). Panel (b) shows that in such a case, the solution does not evolve much: it exhibits weak oscillations around the stationary solution (which is a topological gap soliton). This is especially visible in the periodic variation of the central dimer density.

\end{document}